\documentclass[preprintnumbers,showkeys,amsmath,amssymb]{revtex4}
\usepackage{graphicx}
\usepackage{dcolumn}
\usepackage{bm}
\begin{document}

\preprint{}

\title{Heralded quantum repeater based on the scattering of
photons off single emitters in one-dimensional
waveguides\footnote{published in Ann. Phys. \textbf{378}, 33-46
(2017)}}

\author{Guo-Zhu Song$^{1}$, Mei Zhang$^{1}$, Qing Ai$^{1}$, Guo-Jian Yang$^{1}$,
Ahmed Alsaedi$^{2}$, Aatef Hobiny$^{2}$,
Fu-Guo Deng$^{1,2,}$\footnote{Corresponding author:fgdeng@bnu.edu.cn} }

\address{$^{1}$Department of Physics, Applied Optics Beijing Area Major Laboratory,
Beijing Normal University, Beijing 100875, China\\
$^{2}$NAAM-Research Group, Department of Mathematics, Faculty of
Science, King Abdulaziz University, P.O. Box 80203, Jeddah 21589,
Saudi Arabia}

\begin{abstract}
We propose a heralded quantum repeater based on the scattering of
photons off single emitters in one-dimensional waveguides. We show
the details by implementing nonlocal entanglement generation,
entanglement swapping, and entanglement purification modules with
atoms in waveguides, and discuss the feasibility of the repeater
with currently achievable technology. In our scheme, the faulty
events can be discarded by detecting the polarization of the
photons. That is, our protocols are accomplished with  fidelity of
100\% in principle, which is advantageous for implementing realistic
long-distance quantum communication. Moreover, additional atomic
qubits are not required, but only a single-photon medium. Our scheme
is scalable and attractive since it can be realized in solid-state
quantum systems. With the great progress on controlling
atom-waveguide systems, the repeater may be very useful in quantum
information processing in the future.
\end{abstract}



\maketitle

\section{Introduction}      \label{sec1}


Entanglement plays an important role in quantum communication, such
as quantum key distribution \cite{dis,tri}, quantum secret sharing
\cite{sh}, and quantum secure direct communication
\cite{QSDC1,QSDC2}. However, entangled photon pairs are produced
locally and  inevitably suffer from the noise from optical-fiber
channels when they are transmitted to the parties in quantum
communication, which will decrease the coherence of the photon
systems. In order to exchange private information and avoid an
exponential decay of photons over long distance, the scheme for a
quantum repeater was proposed by Briegel et al. \cite{a} in 1998.
Its main idea is to share the entangled photon pairs in small
segments first, avoiding the exponential decay of photons with the
transmission distance, and then use entanglement swapping \cite{b}
and entanglement purification
\cite{EPP1,EPPsimon,EPPsheng1,DEPP1,DEPP2,DEPP3,DEPP4,HEPP2,HEPPWangGY,DuFFHEPP,dengreview}
to create a long-distance entangled quantum channel.


There are some interesting proposals for implementing a quantum
repeater, by utilizing different physical systems
\cite{e,KL,f,litaorepSR,LiTaoPRA}. For example, in 2001, Duan et al.
\cite{e} suggested an interesting proposal to set up a quantum
repeater with atomic ensembles. In 2006, Klein et al. \cite{KL} put
forward  a robust scheme for quantum repeaters with decoherence-free
subspaces. In 2007, using the two-photon Hong-Ou-Mandel
interferometer, Zhao et al. \cite{f} proposed a robust quantum
repeater protocol. In 2016, Li, Yang, and Deng \cite{LiTaoPRA}
introduced a heralded quantum repeater for quantum communication
network based on quantum dots embedded in optical microcavities,
resorting to effective time-bin encoding. The building blocks of
quantum repeaters are experimentally realized by some research
groups, and remarkable progress has been reported
\cite{re,ma,rk,ab,led}.


In the past decade, the interaction between photons and atoms in
high-quality optical microcavities has become one of the most
important methods for implementing quantum computation and quantum
information processing. Some significant achievements
\cite{hi,gh,fi,nes,se} have been made in photon-atom systems in both
theory and experiment. With strong coupling and high-quality
cavities, they can obtain a high-fidelity quantum computation. In
2005, an interesting proposal \cite{Shen}  was proposed to realize
the coupling between a single quantum emitter and a photon in
one-dimensional (1D) waveguides, which can be considered  as a bad
cavity. In 2007, a similar proposal was presented to realize this
coupling using nanoscale surface plasmons \cite{exp}. In 2015,
S\"{o}llner et al. \cite{sollner1} obtained deterministic
photon-emitter coupling in photonic crystal waveguide in experiment.
In their schemes, the coupling between the emitter and the waveguide
is stronger than the atomic decay rate, but weaker than the
waveguide-loss rate, and the atomic spontaneous emission into the
waveguide becomes the main effect, called the Purcell effect. The
emitter-waveguide systems allow for interesting quantum state
manipulation and quantum information processing, such as
entanglement generation \cite{reso,Kuzyk,TCHliew}, efficient optical
switch \cite{switch}, quantum logic gates
\cite{hypercnot1,renbaocang}, and quantum state transfer
\cite{sca,man,Anpra}. However, with emitter decay and finite
coupling strength, the physical device is restricted to finite $P$
(Purcell factor), so that the scattering of photons off single
emitters may not happen at all. To solve this problem, in $2012$, Li
et al. \cite{sca} proposed a simple scattering setup to realize a
robust-fidelity atom-photon entangling gates, in which the faulty
events can be heralded by detecting the polarization of the photon
pulse.

In this paper, we present a heralded quantum repeater that allows
the nonlocal creation of the entangled state over an arbitrary large distance
with a tolerability of errors. In our scheme, since atoms can
provide long coherence time, we choose a four-level atom as the
emitter. With the scattering of photons off single emitters in 1D waveguides,
the parties in quantum communication can realize nonlocal entanglement
creation against collective noise, entanglement swapping, and
entanglement purification. Moreover, our protocols can turn errors into the detection
of photon polarization, which can be discarded. The prediction of faulty events ensures
that our repeater can be completed with a fidelity of 100\% in principle,
which is advantageous for quantum information processing.

\section{The scattering of photons off single emitters in a 1D waveguide}
\label{basic}

\begin{figure}[!h]
\begin{center}
\includegraphics[width=13 cm,angle=0]{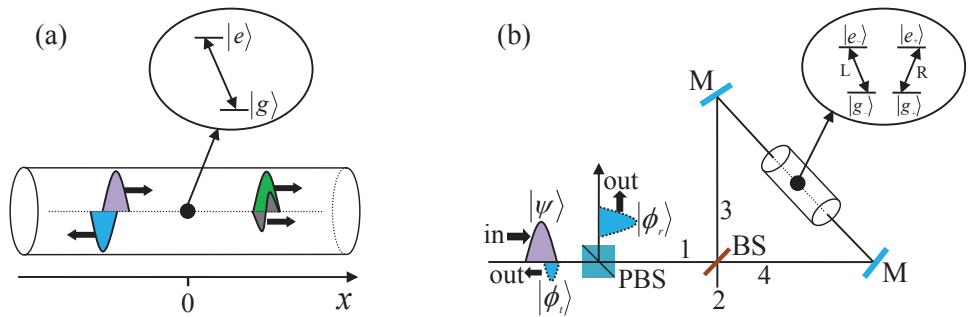}
\caption{  (Color online) (a) The basic structure for a photon mirror in which  a
two-level atom (an emitter marked by the black dot) is coupled to a
1D waveguide (marked by the cylinder). Here the atom has a ground
state $|g\rangle$ and an excited state $|e\rangle$, and
its position is $x=0$. In an ideal situation, an incident photon
(purple, the upper left wave
packet) is fully reflected (blue, the lower left wave packet) when
it resonates with the atom, but there is a transmitted component
(black, the right wave packet) in a practical scattering \cite{exp}. Note
that, when the incident photon is detuned from the emitter , it
goes through the atom with no effect (green, the upper right wave
packet). (b) A heralded setup to realize the scattering between the
photon and emitter in a 1D waveguide \cite{sca}. Different from the emitter in Fig.
1(a), the atom has two degenerate ground states
$|g_{\pm}\rangle$ and two degenerate excited states
$|e_{\pm}\rangle$ coupled to the waveguide (marked by the
cylinder). $PBS$ represents  a polarizing beam splitter
which transmits the horizontal polarized
photon $|H\rangle$ and reflects the vertical polarized photon
$|V\rangle$, BS is a $50:50$ beam splitter, M is a fully reflected
mirror, and the black lines represent the paths of the traveling photon.
} \label{fig1}
\end{center}
\end{figure}

Let us consider a quantum system composed of a single emitter
coupled to electromagnetic modes in a 1D waveguide, as shown in Fig.
\ref{fig1}(a). We first choose a simple two-level atom as the
emitter, consisting of the ground state $|g\rangle$ and the excited
state $|e\rangle$ with the frequency difference $\omega_{a}$. Under
the Jaynes-Cummings model, the Hamiltonian for the interactions
between a set of waveguide modes and a two-level emitter reads
\cite{Shen,exp}:
\begin{eqnarray}     
H=\sum_{k}\!\hbar\omega_{_k}a_{_k}^{\dag}a_{_k}+\frac{1}{2}\hbar\omega_{a}\sigma_{z}
+\sum_{k}\!\hbar
g(a_{_k}^{\dag}\sigma_{_-}\!+a_{_k}\sigma_{_+}),\;\;
\end{eqnarray}
where  $a_{_{k}}$ and $a_{_{k}}^{\dag}$ are the annihilation and creation
operators of the waveguide mode with frequency $\omega_{_{k}}$, respectively.
$\sigma_{z}$, $\sigma_{_{+}}$, and $\sigma_{_{-}}$ are the inversion,
raising, and lowering operators of the two-level atom, respectively.
$g$ is the coupling strength between the atom  and the electromagnetic
modes of the 1D waveguide, assumed to be same for all modes. One can
rewrite the Hamiltonian of the system in real space as \cite{Shen,exp}
\begin{eqnarray}    \label{eqa2} 
H'=\hbar\!\int\!\! dk\, \omega_{_k}a_{_k}^{\dag}a_{_k}
+\hbar g\!\!\int \!\! dk (a_{_k}\sigma_{_+}e^{ikx_{a}}+h.c.)
+\,\hbar(\omega_{a}-\frac{i\gamma'}{2})\sigma_{ee},
\end{eqnarray}
where $x_{a}$ is the position of the atom, $\sigma_{ee}=|e\rangle\langle
e|$, and $\omega_{_k}=c|k|$ ($c$ is the group velocity of
propagating electromagnetic modes and $k$ is its wave vector).
$\gamma'$ is the decay rate of the atom out of the waveguide (e.g.,
the emission into the free space). Because we only care the
interactions of the near-resonant photons with the atom, we could
make the approximation that left- and right-propagating photons form
completely separate quantum fields \cite{Shen}. Under this
approximation, the operator ${a}_{_{k}}$ in Eq. (\ref{eqa2}) can be
replaced by (${a}_{_{k,R}}+{a}_{_{k,L}}$).

To get the reflection and transmission coefficients of single-photon
scattering, we assume that a photon with the energy $E_{k}$ is
propagating from the left. The state of the system is described by \cite{Shen,exp}
\begin{eqnarray}     
|E_{k}\rangle=c_{e}|e,vac\rangle+\int\!\!
dx \Large[\phi_{_{L}}(x)c_{_{L}}^{\dag}(x)
+\phi_{_{R}}(x)c_{_{R}}^{\dag}(x)\Large]|g,vac\rangle,
\label{eqa3}
\end{eqnarray}
where $|vac\rangle$ represents the vacuum state of photons, $c_{e}$ is the
probability amplitude of the atom in the excited state, and
$c_{_{L}}^{\dag}(x)$ ($c_{_{R}}^{\dag}(x)$) is a bosonic operator creating a
left-going (right-going) photon at position $x$. $\phi_{_{R}}(x)$ and $\phi_{_{L}}(x)$ are
the probability amplitudes of right- and left-traveling photon, respectively.
Note that the photon propagates from the left, $\phi_{_{R}}(x)$ and
$\phi_{_{L}}(x)$ could take the forms \cite{Shen,exp}
\begin{eqnarray}     
\phi_{_R}(x)&=& e^{ikx}\theta(-x)+te^{ikx}\theta(x),\nonumber\\
\phi_{_L}(x)&=& re^{-ikx}\theta(-x).
\end{eqnarray}
Here $t$ and $r$ are the transmission and reflection coefficients,
respectively. The Heaviside step function $\theta(x)$ equals 1 when
$x$ is larger than zero and 0 when $x$ is smaller than zero.  By
solving the time-independent Schr\"odinger equation
$H|E_{k}\rangle=E_{k}|E_{k}\rangle$,  one can obtain \cite{Shen,exp}
\begin{eqnarray}     
r&=&-\frac{1}{1+\gamma'/\gamma_{_{1D}}-2i\Delta/\gamma_{_{1D}}}, \nonumber\\
t&=&1+r,
 \label{eqa5}
\end{eqnarray}
where $\Delta=\omega_{_{k}}-\omega_{a}$ is the photon detuning with the
two-level atom, and $\gamma_{_{1D}}=4\pi g^{2}/c$ is the decay rate of
the atom into the waveguide.

Provided that the incident photon resonates with the emitter (i.e.,
$\Delta=0$), one can easily obtain the reflection coefficient
$r=-1/(1+1/P)$, where $P=\gamma_{_{1D}}/{\gamma'}$ is the Purcell
factor. As we know, in the atom-waveguide system, the spontaneous
emission rate $\gamma_{_{1D}}$ into the 1D waveguide can be much
larger than the emission rate $\gamma'$ into all other possible
channels \cite{Shen,exp}. Considering that a high Purcell factor $P$
can be obtained in realistic systems \cite{exp}, one can get the
reflection coefficient $r\approx-1$ for this system in principle. That is, when the
photon is coupled to the emitter, the atom acts as a photon mirror
\cite{Shen}, which puts a $\pi$-phase shift on reflection. However,
when the photon is detuned from the emitter, it transmits
through the atom with no effect.

Let us consider a four-level atom with degenerate ground states
$|g_{\pm}\rangle$ and excited states $|e_{\pm}\rangle$ as the
emitter in a 1D waveguide, as shown in Fig. \ref{fig1}(b). For the
emitter, the transitions of
$|g_{+}\rangle\leftrightarrow|e_{+}\rangle$ and
$|g_{-}\rangle\leftrightarrow|e_{-}\rangle$ are coupled to two
electromagnetic modes $a_{_{k,R}}$ and $a_{_{k,L}}$, with the
absorption (or emission) of right (R) and left (L) circular
polarization photons, respectively. Assuming that the spatial wave
function of the incident photon is $|\psi\rangle$, with the
scattering  properties in the practical situation discussed above,
one can get \cite{sca}
\begin{eqnarray}   
|g_{+}\rangle|\psi\rangle|R\rangle  &\rightarrow |g_{+}\rangle|\phi\rangle|R\rangle,\;\;\;\;\;\;\;
|g_{-}\rangle|\psi\rangle|R\rangle  &\rightarrow |g_{-}\rangle|\psi\rangle|R\rangle,\nonumber\\
|g_{-}\rangle|\psi\rangle|L\rangle  &\rightarrow |g_{-}\rangle|\phi\rangle|L\rangle,\;\;\;\;\;\;\;
|g_{+}\rangle|\psi\rangle|L\rangle  &\rightarrow |g_{+}\rangle|\psi\rangle|L\rangle.
\end{eqnarray}
Here $|\phi\rangle$ is the spatial state of the photon component
left in the waveguide after the scattering process. In general
situation, $|\phi\rangle=|\phi_{t}\rangle+|\phi_{r}\rangle$, where
$|\phi_{t}\rangle=t|\psi\rangle$ and
$|\phi_{r}\rangle=r|\psi\rangle$ refer to the transmitted and
reflected parts of the photon, respectively. When the Purcell factor
$P$ is infinite, $|\phi\rangle$ is normalized. Whereas, if the input
photon is in the horizontal linear-polarization state
$|H\rangle=(|R\rangle+|L\rangle)/\sqrt{2}$, the transformations turn
into \cite{sca}
\begin{eqnarray}      
|g_{+}\rangle|\psi\rangle|H\rangle&\rightarrow
\frac{1}{2}|g_{+}\rangle[(|\phi\rangle+|\psi\rangle)|H\rangle
+(|\phi\rangle-|\psi\rangle)|V\rangle],\nonumber\\
|g_{-}\rangle|\psi\rangle|H\rangle&\rightarrow
\frac{1}{2}|g_{-}\rangle[(|\phi\rangle+|\psi\rangle)|H\rangle
-(|\phi\rangle-|\psi\rangle)|V\rangle],
\label{eqa7}
\end{eqnarray}
where $|V\rangle=(|R\rangle-|L\rangle)/\sqrt{2}$ is the vertical
linear-polarization state. Following the relation in Eq.
(\ref{eqa5}), one gets
$(|\phi\rangle+|\psi\rangle)/2=|\phi_{t}\rangle$ and
$(|\phi\rangle-|\psi\rangle)/2=|\phi_{r}\rangle$ \cite{sca}, and the
transformations in Eq. (\ref{eqa7}) are equivalent to
\begin{eqnarray}      
|g_{+}\rangle|\psi\rangle|H\rangle&
\rightarrow\,|g_{+}\rangle|\phi_{t}\rangle|H\rangle+|g_{+}\rangle|\phi_{r}\rangle|V\rangle,\nonumber\\
|g_{-}\rangle|\psi\rangle|H\rangle&
\rightarrow\,|g_{-}\rangle|\phi_{t}\rangle|H\rangle-|g_{-}\rangle|\phi_{r}\rangle|V\rangle.
\end{eqnarray}
It is interesting that the scattering process generates a
vertical-polarized component. Moreover, for the outgoing photon in
state $|H\rangle$, nothing happens to the emitter, while for the
photon component in state $|V\rangle$, a state-dependent $\pi-$phase
shift occurs on the emitter.

With the principle mentioned above, Li et al. \cite{sca} constructed
a heralded setup to realize the scattering between incident photon
and the emitter in a 1D waveguide, as shown in Fig. \ref{fig1}(b).
The input photon in spatial state $|\psi\rangle$ with $|H\rangle$
(from port $1$) is  split by a 50 : 50 beam splitter (BS) into two
halves that scatter with the atom simultaneously. Then, the
transmitted and reflected components travel back and exit the beam
splitter from port $1$. The corresponding transformations on the
states can be described as follows \cite{sca}:
\begin{eqnarray}\label{eq8}  
\!\!\vert
\Phi_0\rangle&\;\;=&|g_{\pm}\rangle|\psi\rangle|H\rangle^{^{1}}\nonumber\\
\!\!\!\!&\stackrel{ BS}{\longrightarrow}
&\frac{1}{\sqrt{2}}|g_{\pm}\rangle|\psi\rangle|H\rangle^{^{3}}
+\frac{1}{\sqrt{2}}|g_{\pm}\rangle|\psi\rangle|H\rangle^{^{4}}\nonumber\\
\!\!\!\!&\stackrel{ S\!catter}{\longrightarrow}
&\frac{1}{\sqrt{2}}|g_{\pm}\rangle|\phi_{t}\rangle|H\rangle^{^{3}}
+\frac{1}{\sqrt{2}}|g_{\pm}\rangle|\phi_{t}\rangle|H\rangle^{^{4}}
\pm\frac{1}{\sqrt{2}}|g_{\pm}\rangle|\phi_{r}\rangle|V\rangle^{^{3}}
\pm\frac{1}{\sqrt{2}}|g_{\pm}\rangle|\phi_{r}\rangle|V\rangle^{^{4}}\nonumber\\
\!\!\!\!&\stackrel{BS}{\longrightarrow}&|g_{\pm}\rangle|\phi_{t}\rangle
|H\rangle^{^{1}}\pm|g_{\pm}\rangle|\phi_{r}\rangle|V\rangle^{^{1}}.
\end{eqnarray}
Here the superscript $i$ ($i$=1,2,3,4) is the path of the photon, shown in Fig. \ref{fig1}(b).

Note that, due to quantum destructive interference, there is no
photon component coming out from port $2$. Finally, with the help of
$PBS$, discarding the horizontal polarization output from Eq.
(\ref{eq8}) (i.e., the faulty event), one can get the
transformations as follows \cite{sca}:
\begin{eqnarray}          \label{eq9}
|g_{-}\rangle|\psi\rangle|H\rangle &&\rightarrow\; -|g_{-}\rangle|\phi_{r}\rangle|V\rangle,\nonumber\\
|g_{+}\rangle|\psi\rangle|H\rangle &&\rightarrow\; +|g_{+}\rangle|\phi_{r}\rangle|V\rangle.
\end{eqnarray}
Similarly, when the incident photon is in state $|V\rangle$, discarding the faulty event
with vertical polarization output, the transformations are described as follows \cite{sca}:
\begin{eqnarray}          \label{eq10}
|g_{-}\rangle|\psi\rangle|V\rangle &&\rightarrow\; -|g_{-}\rangle|\phi_{r}\rangle|H\rangle,\nonumber\\
|g_{+}\rangle|\psi\rangle|V\rangle &&\rightarrow\; +|g_{+}\rangle|\phi_{r}\rangle|H\rangle.
\end{eqnarray}
As mentioned above, $|\phi_{r}\rangle$ is the spatial wave function
of the reflected photon component after the scattering process. When
$P\rightarrow\infty$, the perfect scattering process leads to
$|\phi_{r}\rangle=-|\psi\rangle$. In the imperfect situation with a
finite $P$, there is always a transmitted part \cite{exp}, we get
$|\phi_{r}\rangle\neq-|\psi\rangle$, the output photon with
unchanged polarization is detected and the corresponding scattering
event fails, which can be discarded. That is, the setup for
realizing the scattering event between incident photon and the
emitter works in a heralded way.

\section{Quantum repeater based on the scattering configuration} \label{sec4}

\subsection{Robust nonlocal entanglement creation against collective noise}
 \label{creation}

\begin{figure}
\begin{center}
\includegraphics[width=8.0cm,angle=0]{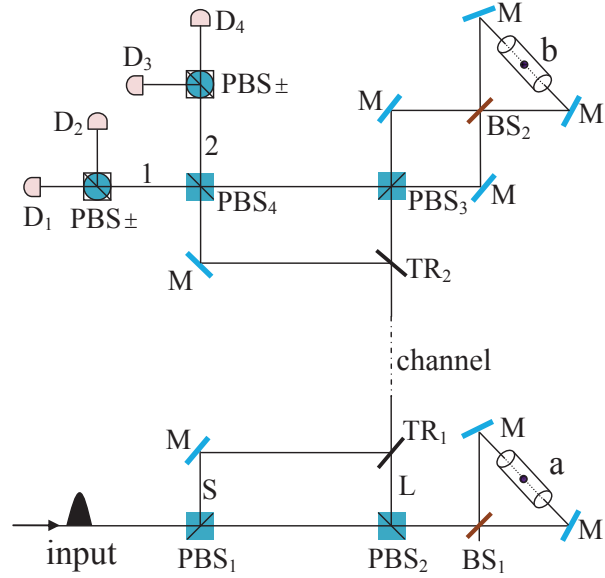}
\caption{ (Color online)  Schematic setup for the creation of  maximally entangled
states on two nonlocal atoms $a$ and $b$ in 1D waveguides.
$PBS\pm$ transmits photons with polarization
$|+\rangle$ and reflects photons with polarization $|-\rangle$,
where $|\pm\rangle=(1/\sqrt{2})(|H\rangle\pm|V\rangle)$. $D_i$
($i=1,2,\cdots,4$) is a photon detector and
$TR_i$ ($i=1,2$) is an optical device which can be controlled exactly as
needed to transmit or reflect a photon.}\label{fig2}
\end{center}
\end{figure}

With the  property of a photon scattering with a four-level atom
coupled to a 1D waveguide,  we can design a robust scheme for the
entanglement creation on two nonlocal stationary atoms $a$ and $b$,
as shown in Fig. \ref{fig2}. Suppose that the single photon medium
and the two stationary atoms in 1D waveguides are initially prepared
in the superposition states
$|\psi_{_{0}}\rangle^{p}=\frac{1}{\sqrt{2}}(|H\rangle+|V\rangle)$
and
$|\varphi_{i}\rangle=\frac{1}{\sqrt{2}}(|0\rangle+|1\rangle)_{i}$
(here $|0\rangle=|g_{-}\rangle,|1\rangle=|g_{+}\rangle$, $i$ = a,
b), respectively, the  state of the system composed of the photon
and the two atoms is
\begin{eqnarray}   
|\Omega_{0}\rangle=\frac{1}{2\sqrt{2}}(|H\rangle+|V\rangle)
\otimes(|0\rangle+|1\rangle)_{a}\otimes(|0\rangle+|1\rangle)_{b}.\;\;
\end{eqnarray}
Our scheme works with the following steps.

First, the $|H\rangle$ and $|V\rangle$ components of the  input
photon are spatially split by a polarizing beam splitter ($PBS$). In
detail, the photon in state $|H\rangle$ passes through both
$PBS_{1}$ and $PBS_{2}$ towards the setup to scatter with atom $a$.
While the component in state $|V\rangle$ is reflected into the other
arm of the interferometer by $PBS_{1}$, and is reflected by $TR_1$
into the channel, having no interaction with atom $a$. After the
scattering process, the part interacting with atom $a$ travels
through $TR_1$ into the channel, but a little later than the other
part. The state of the whole system at the entrance of the channel
becomes $|\Omega_{1}\rangle$. Here
\begin{eqnarray}   
|\Omega_{1}\rangle&=&|\Omega_{_{1S}}\rangle+|\Omega_{_{1L}}\rangle,\nonumber\\
|\Omega_{_{1S}}\rangle&=&\frac{1}{2\sqrt{2}}|V\rangle_{_{S}}
\otimes (|0\rangle+|1\rangle)_{a}
\otimes (|0\rangle+|1\rangle)_{b},\nonumber\\
|\Omega_{_{1L}}\rangle&=&\frac{1}{2\sqrt{2}}|V\rangle_{_{L}}
\otimes (|0\rangle-|1\rangle)_{a}
\otimes (|0\rangle+|1\rangle)_{b},\;\;\;\;\;\;
\end{eqnarray}
where  $|\Omega_{_{1S}}\rangle$ and $|\Omega_{_{1L}}\rangle$ represent
the two parts of the photon going through the short path (S) and the long path (L)
to the channel, respectively.

Second, as the two parts in the channel are near and their
polarization states are both in $|V\rangle$, the influences of the
collective noise in the quantum channel on these two parts are the
same one \cite{PLA,yamamoto,lixihan}, which can be described by
$|V\rangle\;\rightarrow\;\gamma|V\rangle+\delta|H\rangle$, where
$|\gamma|^{2}+|\delta|^{2}=1$. After the photon travels in the long
quantum channel, the state of the whole system at the output
port becomes 
\begin{eqnarray}   
|\Omega_{2}\rangle &=&|\Omega_{_{2S}}\rangle+|\Omega_{_{2L}}\rangle,\nonumber\\
|\Omega_{_{2S}}\rangle &=& \frac{1}{2\sqrt{2}}(\gamma|V\rangle_{_{S}}
+\delta|H\rangle_{_{S}})(|0\rangle+|1\rangle)_{a}(|0\rangle+|1\rangle)_{b},\nonumber\\
|\Omega_{_{2L}}\rangle &=& \frac{1}{2\sqrt{2}}(\gamma|V\rangle_{_{L}}
+\delta|H\rangle_{_{L}})(|0\rangle-|1\rangle)_{a}(|0\rangle+|1\rangle)_{b}.
\end{eqnarray}

Third, getting out from the noisy channel, the early part of the
photon in state $|\Omega_{_{2S}}\rangle$ is reflected by the optical
device $TR_{2}$, while the late part in state
$|\Omega_{_{2L}}\rangle$ transmits through $TR_{2}$ into $PBS_{3}$.
After that, the components in states $|H\rangle$ and $|V\rangle$ of
the late part are split into two halves that scatter with atom $b$
and travel back to $PBS_{3}$ simultaneously. Subsequently, the early
part and the late part are rejoined in $PBS_{4}$, and they are
separated into two paths $1$ and $2$. The state of the whole system
evolves into
\begin{eqnarray}   \label{eq15}
|\Omega_{3}\rangle&=&\frac{1}{2\sqrt{2}}\gamma(|H\rangle+|V\rangle)_{1}
(|0\rangle|0\rangle+|1\rangle|1\rangle)_{ab}
-\frac{1}{2\sqrt{2}}\gamma(|H\rangle-|V\rangle)_{1}(|0\rangle|1\rangle
+|1\rangle|0\rangle)_{ab}\nonumber\\
&&+\frac{1}{2\sqrt{2}}\delta(|H\rangle+|V\rangle)_{2}(|0\rangle|0\rangle
+|1\rangle|1\rangle)_{ab}
+\frac{1}{2\sqrt{2}}\delta(|H\rangle-|V\rangle)_{2}(|0\rangle|1\rangle
+|1\rangle|0\rangle)_{ab}.\;\;\;\;\;\;
\end{eqnarray}

Finally, the two parts in paths $1$ and $2$  go through $PBS\pm$,
and the photon is detected by one of the four single-photon
detectors $D_{1}$, $D_{2}$, $D_{3}$, and $D_{4}$. If the detector
$D_{2}$ or $D_{3}$ clicks, we should put a $\sigma_{x}$ operation on
atom $b$. If the detector $D_{1}$ or $D_{4}$ clicks, nothing needs
to be done. Eventually, the state of the system composed of atoms
$a$ and $b$ collapses to the maximally entangled state
$|\phi^{+}\rangle_{ab}=\frac{1}{\sqrt{2}}(|0\rangle|0\rangle+|1\rangle|1\rangle)_{ab}$.

Note that, for successful events of imperfect processes, i.e., with
finite $P$, the polarization is swapped but
$|\phi_{r}\rangle\neq-|\psi\rangle$ in Eq. (\ref{eq9}) and Eq.
(\ref{eq10}). This causes a problem that the spatial wave functions
in two arms of the interferometer are not matched at $PBS_{4}$. To
overcome the unbalance between the two spatial wave functions, a
waveform corrector ($WFC$) is adopted in one arm of the
interferometer. In fact, the $WFC$ can be realized by a second
scattering module, which is identical to that of Fig. \ref{fig1}(b).
In detail, we make the auxiliary emitter in $WFC$ permanently stay
in $|g_{-}\rangle$, before and after the scattering process, a
quarter wave plate is needed to implement
$|V\rangle$$\leftrightarrow$$|L\rangle$. With the waveform
correctors, the corresponding wave packet is changed from
$|\psi\rangle$ to $|\phi_{r}\rangle$ without entangling with the
auxiliary emitter. The $WFC$ decreases the overall success
probability of the entanglement creation, but not affect the
fidelity in principle.

Our setup for the robust entanglement creation on two nonlocal atoms
has some interesting features. First, the early part and the late
part of the photon in the channel are so near that they suffer from
the same collective noise \cite{PLA,yamamoto,lixihan}, and an
arbitrary qubit error caused by the long noisy channel can be
perfectly settled, i.e., as shown in Eq. ($\ref{eq15}$), the
probability of the entanglement creation doesn't depend on the
values of collective noise parameters $\gamma$ and $\delta$. Second,
the faulty interactions between the photon and two atoms can be
heralded by the detectors $D_{1}$, $D_{2}$, $D_{3}$, and $D_{4}$. In
detail, if none of these detectors clicks, the event of the
entanglement creation fails, which can be discarded. These good
features make our setup have good applications in quantum repeaters
for long-distance quantum communication.

\subsection{Entanglement swapping}
\label{swapping}

\begin{figure*}[tpb]        
\begin{center}
\includegraphics[width=15cm,angle=0]{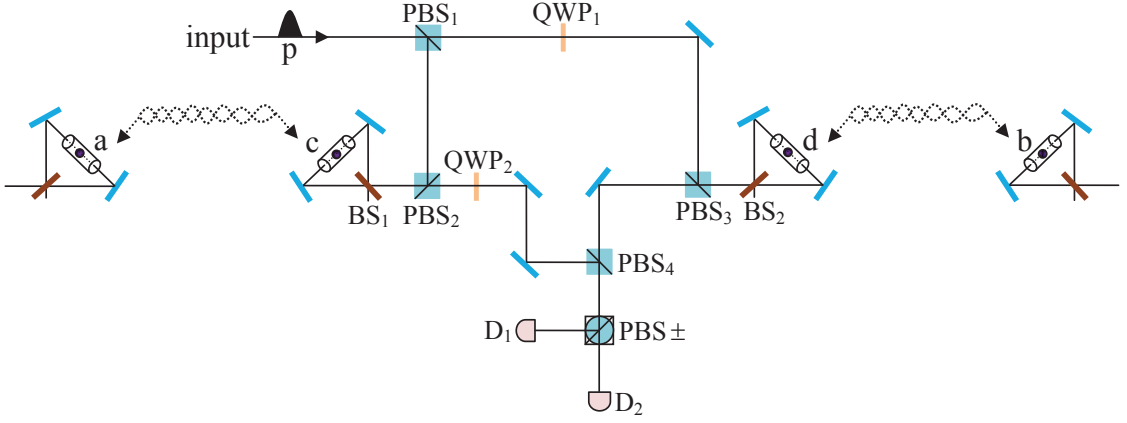}
\caption{ (Color online) Schematic diagram showing the principle of entanglement
swapping. $QWP_{i}$ ($i=1,2$) represents a quarter-wave plate, which
is used to implement the conversion of the photon polarization. }
\label{fig3}
\end{center}
\end{figure*}

The atomic entangled state can be connected to longer communication
distance via local entanglement swapping. Inspired by the recent
work \cite{Sahandarxiv}, we construct the entanglement swapping
scheme, using the scattering of a single photon combined with
measurements on the atoms, as shown in Fig. \ref{fig3}. The two
pairs of nonlocal atoms $ac$ and $bd$ are both initially prepared in
the maximally entangled states
$|\phi^+\rangle_{ac}=\frac{1}{\sqrt{2}}(|0\rangle|0\rangle+|1\rangle|1\rangle)_{ac}$
and
$|\phi^+\rangle_{bd}=\frac{1}{\sqrt{2}}(|0\rangle|0\rangle+|1\rangle|1\rangle)_{bd}$,
respectively. With the Bell-state measurement on local atoms $cd$
and single-qubit operations, the two nonlocal atoms $ab$ can
collapse to the maximally entangled state
$|\phi^+\rangle_{ab}=\frac{1}{\sqrt{2}}(|0\rangle|0\rangle+|1\rangle|1\rangle)_{ab}$,
which indicates that the nonlocal entanglement for a longer
communication is realized. The principle of quantum swapping is
shown in Fig. \ref{fig3}, and the details are described as follows.

First, suppose that the input photon $p$ is prepared in the
superposition state $|\psi_{_{0}}\rangle^{p}=\frac{1}{\sqrt{2}}(|H\rangle+|V\rangle)$,
and the initial state of the whole system composed of photon $p$  and
the four atoms $acbd$ is $|\Psi_{0}\rangle$. Here,
\begin{eqnarray}   
|\Psi_{0}\rangle&=&\frac{1}{2\sqrt{2}}(|H\rangle\!+
\!|V\rangle)\otimes(|0\rangle|0\rangle\!+\!|1\rangle|1\rangle)_{ac}
\otimes(|0\rangle|0\rangle \!+\!|1\rangle|1\rangle)_{bd}.
\end{eqnarray}
The injecting photon $p$ passes through $PBS_{1}$, which transmits the
photon in state $|H\rangle$ and reflects  the photon in state
$|V\rangle$. The photon in state $|V\rangle$ is reflected by
$PBS_{1}$ and $PBS_{2}$ into the scattering setup composed of atom
$c$, while the other part in state $|H\rangle$ goes through
$QWP_{1}$ and is reflected by $PBS_{3}$ into the scattering setup to scatter with
atom $d$. Then, the two parts of the photon $p$ are
rejoined in $PBS_{4}$. After that,
the state of the whole system is changed from $|\Psi_{0}\rangle$ to $|\Psi_{1}\rangle$. Here,
\begin{eqnarray}      
|\Psi_{1}\rangle&=&\frac{(|H\rangle+|V\rangle)}{4\sqrt{2}}\big[(|00\rangle-|11\rangle)_{cd}
\otimes(|00\rangle+|11\rangle)_{ab}
+(|00\rangle+|11\rangle)_{cd}\otimes(|00\rangle-|11\rangle)_{ab}\big]\nonumber\\
&&-\frac{(|H\rangle-|V\rangle)}{4\sqrt{2}}\big[\!(|01\rangle-|10\rangle)_{cd}
\otimes(|01\rangle+|10\rangle)_{ab}
+(|01\rangle+|10\rangle)_{cd}\otimes(|01\rangle-|10\rangle)_{ab}\big].
\end{eqnarray}

Second, a Hadamard operation $H_{a}$ (e.g., using a $\pi/2$ microwave pulse or optical pulse
\cite{Berezovsky,Press}) is performed on the two local atoms $c$ and $d$
in the waveguides, respectively. Then, the state of the whole system becomes
\begin{eqnarray}       
|\Psi_{2}\rangle&=&\frac{(|H\rangle+|V\rangle)}{4\sqrt{2}}\big[(|01\rangle+|10\rangle)_{cd}
\otimes(|00\rangle+|11\rangle)_{ab}
+(|00\rangle+|11\rangle)_{cd}\otimes(|00\rangle-|11\rangle)_{ab}\big]\nonumber\\
&&+\frac{(|H\rangle-|V\rangle)}{4\sqrt{2}}\big[\!(|01\rangle-|10\rangle)_{cd}
\otimes(|01\rangle+|10\rangle)_{ab}
-(|00\rangle-|11\rangle)_{cd}\otimes(|01\rangle-|10\rangle)_{ab}\big].
\end{eqnarray}
Then, the photon $p$ travels through $PBS\pm$ and is detected by single-photon detectors.
Meanwhile, the state of atom $c$ ($d$) is measured by external classical field.

Third, with the outcomes of the detectors for photon $p$ and the
measurements on atoms $cd$, one can see that the four Bell states of
the atoms $a$ and $b$ are completely distinguished. Finally, the
parties can perform corresponding operations (see Table
\ref{tabone}) on atom $a$ to complete the quantum swapping. After
that, the state of the two nonlocal atoms $a$ and $b$ in a longer
distance collapses to the maximally entangled state
$|\phi^{+}\rangle_{ab}=\frac{1}{\sqrt{2}}(|0\rangle|0\rangle+|1\rangle|1\rangle)_{ab}$.

It is important to note that the wrong interactions between photon
and atoms are heralded by the photon detectors in our protocol. In
detail, if neither of the detectors $D_{1}$ and $D_{2}$ clicks, the
interactions between photon and two atoms in 1D waveguides are
faulty, which could be discarded. Therefore, with the prediction of
the faulty events, the parties can obtain a high-fidelity nonlocal
atomic entangled state in a longer distance.

\begin{table} [h]        
\centering
\caption{The operations on atom $a$ corresponding to the outcomes
of the photon detectors and the states of atoms $cd$.}
\begin{tabular}{ccc|c}
\hline    \hline
$\;\;$Photon click$\;\;\;\;\;\;$  &  Atom $c$    $\;\;\;\;\;\;$        &    Atom $d$
   $\;\;$    &    $\;$  Operations on atom $a$  $\;$  \\
\hline
$\;\;$$D_{1}$   $\;\;\;\;\;\;$    & $|0\rangle(|1\rangle)$$\;\;\;\;\;\;\;\;$
  &  $|1\rangle(|0\rangle)$  $\;\;$   &$\;\;$       $I$                          $\;\;$  \\
$\;\;$$D_{1}$   $\;\;\;\;\;\;$    & $|0\rangle(|1\rangle)$$\;\;\;\;\;\;\;\;$
  &  $|0\rangle(|1\rangle)$  $\;\;$   &$\;\;$       $\sigma_{z}$                 $\;\;$  \\
$\;\;$$D_{2}$   $\;\;\;\;\;\;$    & $|0\rangle(|1\rangle)$$\;\;\;\;\;\;\;\;$
 &  $|1\rangle(|0\rangle)$  $\;\;$   &$\;\;$       $\sigma_{x}$                 $\;\;$  \\
$\;\;$$D_{2}$   $\;\;\;\;\;\;$    & $|0\rangle(|1\rangle)$$\;\;\;\;\;\;\;\;$
 &  $|0\rangle(|1\rangle)$  $\;\;$   &$\;\;$       $\sigma_{z}\sigma_{x}$       $\;\;$  \\
\hline  \hline
\end{tabular}  \label{tabone}
\end{table}

\subsection{Entanglement purification}
\label{sec43}

\begin{figure*}[tpb]  
\begin{center}
\includegraphics[width=12cm,angle=0]{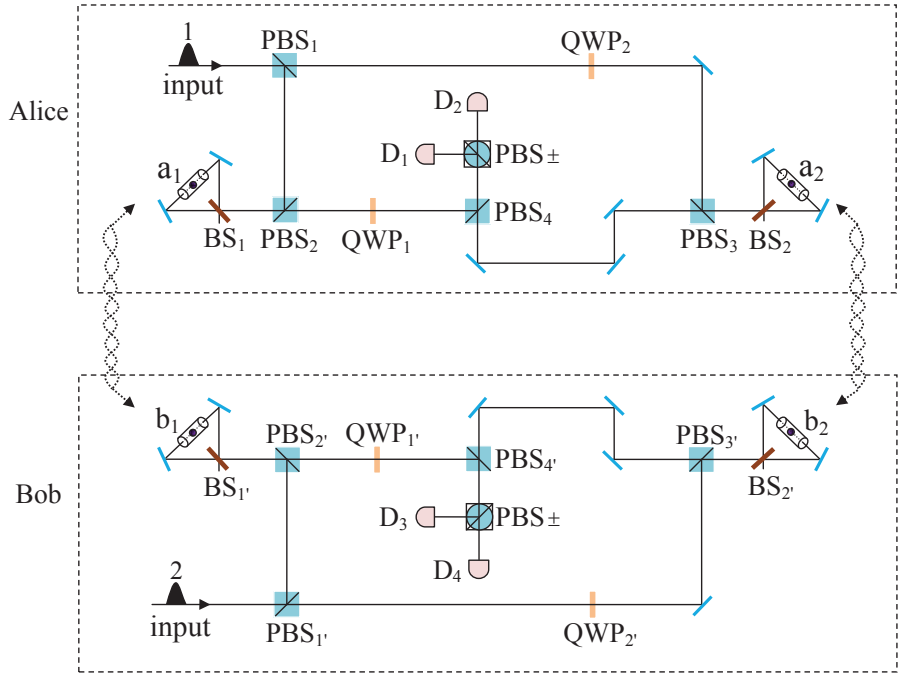}
\caption{(Color online) Schematic setup showing the principle of the
atomic entanglement purification protocol based on the scattering of
photons off single emitters.}     \label{fig4}
\end{center}
\end{figure*}

In section \ref{creation} and section \ref{swapping}, we just talk
about the influence of noise on flying photons in long quantum
channel. In the practical situation, the errors also occur in
stationary atoms embedded in 1D waveguides, which will decrease the
entanglement of the nonlocal two-atom systems. Using entanglement
purification
\cite{EPP1,EPPsimon,EPPsheng1,DEPP1,DEPP2,DEPP3,DEPP4,HEPP2,HEPPWangGY,DuFFHEPP,dengreview},
we can distill some high-fidelity maximally entangled states from a
mixed entangled state ensemble. Now, we start to explain the
principle of our purification protocol for nonlocal atomic entangled
states, assisted by the scattering of photons off single atoms in 1D
waveguides, as shown in Fig. \ref{fig4}.

Suppose that the initial mixed state shared by two remote parties,
say Alice and Bob, can be written as
\begin{eqnarray}   
\rho_{ab}=F|\phi^{+}\rangle_{ab}\langle\phi^{+}|+(1-F)|\psi^{+}\rangle_{ab}\langle\psi^{+}|,
\end{eqnarray}
where $|\psi^{+}\rangle_{ab}=\frac{1}{\sqrt{2}}(|0\rangle|1\rangle
+|1\rangle|0\rangle)_{ab}$. The subscripts $a$ and $b$ represent the
single atoms in 1D waveguides owned by Alice and Bob, respectively.
$F$ is the initial fidelity of the state $|\phi^{+}\rangle$. By
selecting two pairs of nonlocal entangled two-atom systems, the four
atoms are in the states
$|\phi^{+}\rangle_{a_{1}b_{1}}|\phi^{+}\rangle_{a_{2}b_{2}}$ with
the probability of $F^{2}$,
$|\phi^{+}\rangle_{a_{1}b_{1}}|\psi^{+}\rangle_{a_{2}b_{2}}$ and
$|\psi^{+}\rangle_{a_{1}b_{1}}|\phi^{+}\rangle_{a_{2}b_{2}}$ with a
probability of $F(1-F)$, and
$|\psi^{+}\rangle_{a_{1}b_{1}}|\psi^{+}\rangle_{a_{2}b_{2}}$ with a
probability of $(1-F)^{2}$, respectively. Our entanglement
purification protocol for nonlocal entangled atom pairs works with
the following steps.

First, both Alice and Bob prepare an optical pulse in the
superposition state $\frac{1}{\sqrt{2}}(|H\rangle+|V\rangle)$ and
let them pass through the equipments shown in Fig.
\ref{fig4}. Here, we choose the case $|\phi^{+}\rangle_{a_{1}b_{1}}|\phi^{+}\rangle_{a_{2}b_{2}}$
to illustrate the principle. To simplify the discussion, we just discuss the
interactions in Alice, and Bob need complete the same process simultaneously.
For Alice, the $|H\rangle$ and $|V\rangle$
components of the input photon 1 are spatially split by $PBS_{1}$.
In detail, the component in $|V\rangle$ is reflected by both
$PBS_{1}$ and $PBS_{2}$ to the scattering setup composed of atom $a_{1}$,
whereas the component in $|H\rangle$ goes through $QWP_{2}$ and is
reflected by $PBS_{3}$ to the scattering setup containing atom $a_{1}$. After that,
the state of the whole system is changed from $|\Phi_{0}\rangle$ to $|\Phi_{1}\rangle$, where
\begin{eqnarray}           
|\Phi_{0}\rangle&=&\frac{1}{2}(|H\rangle+|V\rangle)_{1}
\,\,(|H\rangle+|V\rangle)_{2}
\,\,|\phi^{+}\rangle_{a_{1}b_{1}}|\phi^{+}\rangle_{a_{2}b_{2}},\nonumber\\
|\Phi_{1}\rangle&=&\frac{1}{4}|H\rangle_{1}
\,(|H\rangle+|V\rangle)_{2}\otimes(0000-0011+1100-1111)_{a_{1}b_{1}a_{2}b_{2}}\nonumber\\
&&+\frac{1}{4}|V\rangle_{1}\,(|H\rangle+|V\rangle)_{2}
\otimes(0000+0011-1100-1111)_{a_{1}b_{1}a_{2}b_{2}}.
\end{eqnarray}

Second, the two parts of photon 1 are rejoined at $PBS_{4}$ and
travel through a $PBS\pm$. Meanwhile, the photon 2 at Bob's place
has the same process as photon 1 in Alice simultaneously. After the
above interactions, the state of the whole system collapses into
$|\Phi_{2}\rangle$. Here
\begin{eqnarray}           
|\Phi_{2}\rangle&=&\frac{1}{4}(|H\rangle\!+\!|V\rangle)_{1}(|H\rangle+|V\rangle)_{2}
\,(|0000\rangle\!+\!|1111\rangle)_{a_{1}b_{1}a_{2}b_{2}}\nonumber\\
&&+\frac{1}{4}(|H\rangle-|V\rangle)_{1}(|H\rangle-|V\rangle)_{2}
\,(|0011\rangle+|1100\rangle)_{a_{1}b_{1}a_{2}b_{2}}.
\end{eqnarray}
Finally, photon 1 and photon 2 are probed by single-photon detectors.

Similarly, the evolution of the other three cases can be described as follows:
\begin{eqnarray}           
\!\!\!\!&&\frac{1}{2}(|H\rangle+|V\rangle)_{1}(|H\rangle+|V\rangle)_{2}
\,|\phi^{+}\rangle_{a_{1}b_{1}}|\psi^{+}\rangle_{a_{2}b_{2}}\nonumber\\
&&\rightarrow-\frac{1}{4}(|H\rangle+|V\rangle)_{1}(|H\rangle-|V\rangle)_{2}
\,(|0001\rangle+|1110\rangle)_{a_{1}b_{1}a_{2}b_{2}}\nonumber\\
&&\;\;\;\;\;-\frac{1}{4}(|H\rangle-|V\rangle)_{1}\!(|H\rangle+|V\rangle)_{2}
\,(|0010\rangle+|1101\rangle)_{a_{1}b_{1}a_{2}b_{2}},
\end{eqnarray}
\begin{eqnarray}           
\!\!\!\!&&\frac{1}{2}(|H\rangle+|V\rangle)_{1}(|H\rangle+|V\rangle)_{2}
\,|\psi^{+}\rangle_{a_{1}b_{1}}|\phi^{+}\rangle_{a_{2}b_{2}}\nonumber\\
&&\rightarrow\frac{1}{4}(|H\rangle+|V\rangle)_{1}(|H\rangle-|V\rangle)_{2}
\,(|0100\rangle+|1011\rangle)_{a_{1}b_{1}a_{2}b_{2}}\nonumber\\
&&\;\;\;\;\;+\frac{1}{4}(|H\rangle-|V\rangle)_{1}(|H\rangle+|V\rangle)_{2}
\,(|0111\rangle+|1000\rangle)_{a_{1}b_{1}a_{2}b_{2}},
\end{eqnarray}
and
\begin{eqnarray}           
\!\!\!\!&&\frac{1}{2}(|H\rangle+|V\rangle)_{1}(|H\rangle+|V\rangle)_{2}
\,|\psi^{+}\rangle_{a_{1}b_{1}}|\psi^{+}\rangle_{a_{2}b_{2}}\nonumber\\
&&\rightarrow-\frac{1}{4}(|H\rangle+|V\rangle)_{1}(|H\rangle+|V\rangle)_{2}
\,(|0101\rangle+|1010\rangle)_{a_{1}b_{1}a_{2}b_{2}}\nonumber\\
&&\;\;\;\;\;-\frac{1}{4}(|H\rangle-|V\rangle)_{1}(|H\rangle-|V\rangle)_{2}
\,(|0110\rangle+|1001\rangle)_{a_{1}b_{1}a_{2}b_{2}}.
\end{eqnarray}
The measurement results of all cases are shown in Table \ref{tab2}.
With the outcomes of four detectors, we can distill
$|\phi^{+}\rangle_{a_{1}b_{1}}|\phi^{+}\rangle_{a_{2}b_{2}}$ and
$|\psi^{+}\rangle_{a_{1}b_{1}}|\psi^{+}\rangle_{a_{2}b_{2}}$ from
the four cases mentioned above.

\begin{table} [h]              \label{tab2}
\centering \caption{The results of the four single-photon detectors
corresponding to the initial entangled states of the four atoms.}
\begin{tabular}{cc|c}
\hline    \hline
$\;\;\;$ Initial   &$\;\;\;\;\;\;$    states  $\;\;\;\;\;\;\;\;\;\;$
 &$\;\;\;\;\;\;\;$ Photons $\;\;$ measurement  $\;\;\;\;$  \\
\hline
$\;\;\;$($a_{1}b_{1}$)     $\;$    &$\;$ ($a_{2}b_{2}$)   $\;\;$  &$\;\;$
 Detector $\;\;$ click  $\;\;\;\;$     \\
\hline
$\;\;\;$$|\phi^{+}\rangle$ $\;$    &$\;$ $|\phi^{+}\rangle$$\;\;$
&$\;\;$$D_{2}D_{4}$ $\;\;\;\;$or $\;\;\;\;$$D_{1}D_{3}$  \\
$\;\;\;$$|\phi^{+}\rangle$ $\;$    &$\;$ $|\psi^{+}\rangle$$\;\;$
&$\;\;$$D_{2}D_{3}$ $\;\;\;\;$or $\;\;\;\;$$D_{1}D_{4}$  \\
$\;\;\;$$|\psi^{+}\rangle$ $\;$    &$\;$ $|\phi^{+}\rangle$$\;\;$
&$\;\;$$D_{2}D_{3}$ $\;\;\;\;$or $\;\;\;\;$$D_{1}D_{4}$  \\
$\;\;\;$$|\psi^{+}\rangle$ $\;$    &$\;$ $|\psi^{+}\rangle$$\;\;$
&$\;\;$$D_{2}D_{4}$ $\;\;\;\;$or $\;\;\;\;$$D_{1}D_{3}$  \\
\hline  \hline
\end{tabular}  \label{tab2}
\end{table}

Third, to recover the entangled states of atoms $a_{1}$ and $b_{1}$,
Alice and Bob should perform a Hadamard operation $H_{a}$ on
the two nonlocal atoms $a_{2}$ and $b_{2}$ in the waveguides, respectively. Then,
Alice and Bob measure the states of the two atoms $a_{2}$ and $b_{2}$, and
compare their results with the help of classical communication. If
the results are the same ones, nothing needs to be done; otherwise,
a $\sigma_{z}$ operation needs to be put on atom $a_{1}$. From Table \ref{tab2},
one can see that there are two cases in the reserved entangled pairs $a_{1}$ and $b_{1}$.
One is $|\phi^{+}\rangle_{a_{1}b_{1}}$ with a probability of $F^{2}$, and
the other one is $|\psi^{+}\rangle_{a_{1}b_{1}}$ with a probability of $(1-F)^{2}$.
Therefore, in the filtered states, the probability of $|\phi^{+}\rangle_{a_{1}b_{1}}$ is
$F^{'}=\frac{F^{2}}{F^{2}+(1-F)^{2}}$.
That is, after the purification process, the fidelity of $|\phi^{+}\rangle_{a_{1}b_{1}}$
becomes $F^{'}$. When $F>\frac{1}{2}$, one can easily get $F^{'}>F$.

\section{Discussion and summary}\label{sec6}

\begin{figure}[!h]  
\begin{center}
\includegraphics[width=14cm,angle=0]{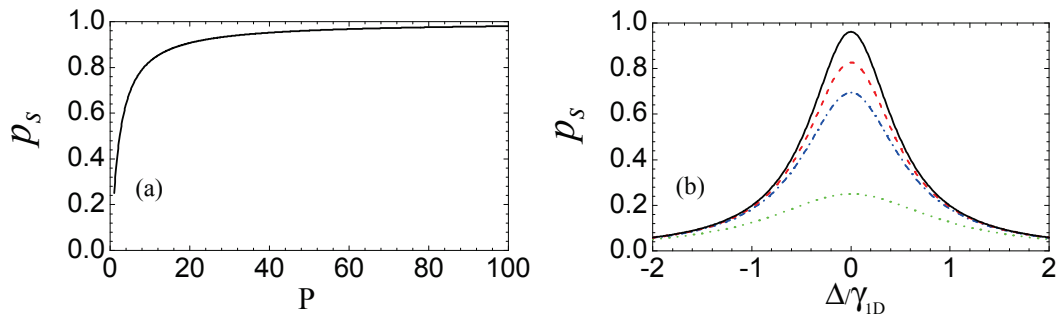}
\caption{(Color online) The success probability $p_{s}$ of the
scattering process vs the Purcell factor $P$ and the detuning
parameter $\Delta/\gamma_{_{1D}}$. (a) The success probability
$p_{s}$ vs the Purcell factor $P$ when the detuning $\Delta=0$. (b)
The success probability $p_{s}$ vs the parameters
$\Delta/\gamma_{_{1D}}$. The dotted (green), dashed-dotted (blue),
dashed (red), and solid (black) lines correspond to $P$=1, $P$=5,
$P$=10, $P$=50, respectively.}      \label{fig5}
\end{center}
\end{figure}

We have proposed a heralded scheme for quantum repeater, including
robust nonlocal entanglement creation against collective noise,
entanglement swapping, and entanglement purification modules. The
key element in our protocol is the scattering process between
photons and atoms in 1D waveguides.  In the following section, we
will discuss the performance of our quantum repeater under practical
conditions, defining $p_{s}=|\langle\psi|\phi_{r}\rangle|^{2}$ as
the success probability of the scattering event in the heralded
protocol, as shown in Fig. \ref{fig1}(b). Here $|\psi\rangle$ and
$|\phi_{r}\rangle$ are the spatial wave functions of the incident
photon and the reflected photon component after the scattering
event, respectively. For perfect scattering event, i.e., with
$P\rightarrow\infty$, $|\phi_{r}\rangle=-|\psi\rangle$, and the
success probability $p_{s}$ is $100\%$. Whereas, in realistic
situations, with finite $P$, $|\phi_{r}\rangle\neq-|\psi\rangle$, it
includes two cases: one is the successful event of imperfect
processes, where the polarization of output photon is changed but
$|\phi_{r}\rangle=r|\psi\rangle$ ($|r|<1$); the other one is that
the scattering event between atoms and photons doesn't happen at
all. For the latter case, the output photons with unchanged
polarization are detected at the entrance, and the corresponding
quantum computation is discarded. Assuming that the linear optical
elements are perfect in our protocols, the heralded mechanism
ensures that the faulty events cannot influence the fidelity of our
scheme, but decrease the efficiency, because the success probability
$p_{s}$ is determined by the quality of the atom-waveguide systems.

\begin{figure}
\begin{center}
\includegraphics[width=8cm,angle=0]{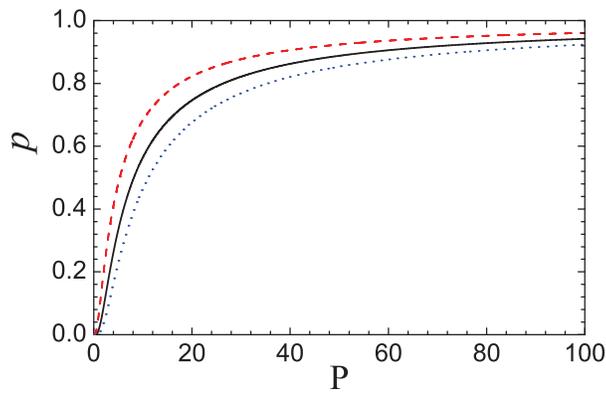}
\caption{ (Color online) The success probabilities of our
entanglement creation  (solid line, black), entanglement swapping
(dashed line, red), entanglement purification (dotted line, blue)
protocols vs the Purcell factor $P$. Here, the detuning parameter is
$\Delta=0$.}      \label{fig6}
\end{center}
\end{figure}

As mentioned above, a high Purcell factor is needed in our scheme,
which can effectively improve the performance of our protocols. In
the last decade, great progress has been made in the
emitter-waveguide systems in both theory and experiment. In 2005,
Vlasov et al. \cite{Vlasov2005} experimentally demonstrated that a
Purcell factor approaching 60 can be observed in low-loss silicon
photonic crystal waveguides. In 2006, Chang et al. \cite{ChangDE}
presented a scheme that a dipole emitter is coupled to a nanowire or
a metallic nanotip, in which a Purcell factor $P=5.2\times10^2$ is
theoretically obtained for a silver nanowire. Subsequently, using
the surface plasmons of a conducting nanowire, Chang et al.
\cite{ChangPRB} proposed a method to obtain an effective Purcell
factor, which can reach $10^3$ in realistic systems in principle. In
addition, short waveguide lengths of only 10 to 20 unit cells were
theoretically found by Manga Rao and Hughes \cite{MangaRao} to
produce a very large Purcell factor in 2007, and the experimental
progress on short photonic crystal waveguides was reported by
Dewhurst et al. \cite{Dewhurst2010} and Hoang et al.
\cite{Hoang2012}. Later, based on subwavelength confinement of
optical fields near metallic nanostructures, Akimov et al.
\cite{Akimov} demonstrated a broadband approach for manipulating
photon-emitter interactions. In 2008, photonic crystal waveguides
were exploited by Hansen et al. \cite{Hansen} to enable single
quantum dots to exhibit nearly perfect spontaneous emission into the
guided modes ($\gamma_{_{1D}}\gg\gamma'$), where the light-matter
coupling strength is largely enhanced. In 2010, a Purcell factor of
$P=5.2$ was experimentally observed for single quantum dots coupled
to a photonic crystal waveguide \cite{Thyr2010}. In 2012, Goban
\cite{GobanKimble} reported the experimental implementation of a
state-insensitive, compensated optical trap for single Cs atoms,
which provides the precise atomic spectroscopy near dielectric
surfaces. Moreover, in 2013, Hung et al. \cite{Hung} proposed a
protocol that  one atom trapped in single nanobeam structure could
provide a resonant probe with transmission $|t|^{2}\leq10^{-2}$ in
theory. A similar scheme was realized in experiment by Goban et al.
\cite{Goban} in 2014. Recently, due to the coupling of a single
emitter to a dielectric slot waveguide,  a high Purcell factor
$P=31$ was also observed by Kolchin et al. \cite{Kolchin} in
experiment.

As illustrated in section \ref{basic}, we can obtain the reflection
coefficient for an incident photon scattering with an atom in the 1D
waveguide. On resonance, $r=-1/(1+1/P)$, and we get the relation
between the success probability $p_{s}$ (i.e., $|r|^{2}$) and the
Purcell factor $P$, as shown in Fig. \ref{fig5}(a). Moreover, the
scattering quality is also influenced by the nonzero photonic
detuning, and the details are described in Fig. \ref{fig5}(b), where
$p_{s}$ is plotted as a function of the detuning parameter
$\Delta/\gamma_{_{1D}}$. From Fig. 5, one can see that the success
probability $p_{s}$ would exceed $90\%$ on the condition that the
Purcell factor $P\geq50$ and the detuning parameter
$\Delta/\gamma_{_{1D}}\leq0.13$, which could be achievable in
realistic systems. For instance, when we choose $P=100$ and
$\Delta=0.1\gamma_{_{1D}}$ for atom-waveguide systems, the success
probability $p_{s}$ of the scattering process in Fig. \ref{fig1}(b)
can reach $94.33\%$. In our protocols, the faulty events between
photons and atomic qubits can be heralded by single-photon
detectors. However, the imperfection coming from photon loss is
still an inevitable problem in our scheme. The photon loss is caused
by various drawbacks, such as the fiber absorption, the imperfection
of 1D waveguides, and the inefficiency of the single-photon
detector. In fact, if optical losses appear in the prediction of
faulty events, the fidelities of our protocols cannot be unity as
faulty scattering events will not always be detected. Recently, many
proposals have been presented to solve the problems of photon loss
\cite{Gisin,Kocsis,Osorio,Wangtiejun}.

Provided that the linear optical elements are perfect in our scheme,
due to the heralded mechanism, only when no faulty scattering events
are detected in our protocols, can we obtain the quantum repeater
successfully. Here, the times of the basic scattering event (Fig.
1(b)) occurred in our entanglement creation, swapping, and
purification protocols are three, two, and four, respectively. We
calculate the success probabilities of our entanglement creation,
swapping, and purification protocols as a function of the Purcell
factor $P$, as shown in Fig. \ref{fig6}. Recently, Arcari et al.
\cite{Arcari2014} exploited the photonic crystal waveguide to
experimentally realize near-unity coupling efficiency of  a quantum
dot to the  waveguide mode. In their experiment, the decay rate of
the quantum dot into the waveguide is $\gamma_{_{1D}}=6.182$ GHz,
and the decay rate of the quantum emitter into free space and all
other modes is $\gamma'=98$ MHz, which indicates that the Purcell
factor $P=63.1$ can be implemented. The resonance frequency of the
quantum dot is $\omega_{a}=2\times10^{6}$ GHz. With these
experimental parameters mentioned above, we can obtain the success
probabilities of our entanglement creation, swapping, and
purification protocols are $p_{1}=91.00\%$, $p_{2}=93.90\%$, and
$p_{3}=88.18\%$, respectively, when the detuning parameter is
$\Delta=0$. With the remarkable progress in photonic nanostructures
\cite{Lodahl2015}, our protocols for the heralded quantum repeater
may be experimentally feasible in the near future.

Compared with other schemes, the protocols we present for the heralded
quantum repeater have some interesting features. First,
the faulty events caused by frequency mismatches, weak coupling, atomic
decay into free space, or finite bandwidth of the incident photon can be
turned into detection of the output photon polarization, and that just
affects the efficiency of our protocols, not the fidelity. In other words,
our scheme either succeeds with a perfect fidelity or fails in
a heralded way, which is an   advantageous feature for quantum
communication. Second, our scheme focuses on 1D waveguides, in which the modes
can be highly dispersive. In a waveguide, the quantum emitter can efficiently couple
single photon to the propagating modes over a wide bandwidth, which provides an
alternative to the high-Q cavity case for enhancing light-matter interaction.
Third, motivated by recent experimental progress, our scheme for the
heralded quantum repeater is feasible in some other quantum
systems, such as superconducting quantum circuit coupled to
transmission lines \cite{Loo}, quantum dot
embedded in a nanowire \cite{Munsch}, and photonic crystal waveguide
with quantum dots \cite{Lodahl2015}.

In summary, we have proposed a heralded quantum repeater based on the
scattering of photons off single emitters in 1D waveguides. The
information of the entangled states is encoded on four-level atoms
embedded in 1D waveguides. As our protocols can transform faulty
events into the detection of photon polarization, we present
a different way for constructing quantum repeaters in solid-state quantum
systems. With the significant progress on manipulating
atom-waveguide systems, our quantum repeater may be very useful for
quantum communication in the future.

\section*{Acknowledgments}

This work was supported by the National Natural Science Foundation
of China under Grants No. 11674033,   No. 11475021,  No. 11505007,
and No. 11474026,  the Fundamental Research Funds for the Central
Universities under Grant No. 2015KJJCA01, the National Key Basic
Research Program of China under Grant No. 2013CB922000, the Youth
Scholars Program of Beijing Normal University under Grant No.
2014NT28, and the Open Research Fund Program of the State Key
Laboratory of Low-Dimensional Quantum Physics, Tsinghua University
Grant No. KF201502.


\end{document}